# THE EFFICIENCY OF QUANTUM TOMOGRAPHY BASED ON PHOTON DETECTION

*Yu. I. Bogdanov[1,3,4], S. P. Kulik[2]*

1 Institute of Physics and Technology, Russian Academy of Sciences, 117218, Moscow, Russia.
2 Faculty of Physics, Moscow M.V.Lomonosov State University, 119992, Moscow, Russia
3 National Research Nuclear University 'MEPHI', 115409 Moscow, Russia
4 National Research University of Electronic Technology MIET, 124498 Moscow, Russia

We propose a general methodology for efficient statistical reconstruction of a quantum state through collection and analysis of photon counting statistics. Our approach includes both strict quantitative criteria for adequacy and completeness of the statistical inverse problem, as well as a simple and reliable method for evaluating errors in the reconstructed state and approximation of a quantum state by means of the reduced finite-dimensional model.

**Introduction**

One of the cornerstones of quantum mechanics is the Niels Bohr's principle of complementarity. According to the principle "evidence obtained under different experimental conditions cannot be comprehended within a single picture, but must be regarded as complementary in the sense that only the totality of the phenomena exhausts the possible information about the objects" [1]. Importance of the principle of complementarity was best described by Wolfgang Pauli claiming that «we might call modern quantum theory as" The Theory of Complementarity "(in analogy with the terminology" Theory of Relativity ")» [2].

A formal description of a quantum state, based on the concepts of the state vector and the density matrix is not limited to a single probability distribution. From a statistical point of view, the quantum state can be regarded as a natural generalization of the concept of a probability distribution. Under the principle of complementarity, an experimental study of the quantum state must be based on the measurement of a combination of mutually complementary distributions. For example, an experimental study of the quantum system will be more complete if the data obtained in the study of a quantum ensemble in the coordinate space is complemented by studying the same ensemble in a canonically conjugate (momentum) space.

A well-known Pauli problem is actually the one to reconstruct the psi function on the basis of the coordinate and momentum distributions [2]. Through the development of homodyne technology in modern quantum optics, the problem of measuring a quantum state has moved from a purely theoretical area to the actual implementation in quantum information technology and in a wider sense than originally anticipated by Pauli. Moreover at the moment it serves as a more or less standard tool for estimating quality of quantum states prepared in experiment. It was found that by rotating the phase of the local oscillator (LO), one can study various quadrature observables of the electric field [3,4]. In this case, the coordinate and momentum are just two special cases corresponding to the phases 0 and $\pi/2$, respectively [5,6].

Let us consider the simplified scheme shown in Figure 1 to describe some of the basic concepts and facts, as well as to introduce the notation that we will need in the future. The input 1 of a beamsplitter is served by a coherent mode from the local oscillator (LO) $|\alpha_1\rangle$ set by the complex amplitude $\alpha_1 = |\alpha_1|\exp(i\theta_1)$. Through the setting of $m$ different local oscillator phases $\theta_1 = 0, \pi/m, ..., \pi(m-1)/m$, one can derive $m$ different distributions from a set of mutually complementary distributions.

The input 2 of a beamsplitter served by the quantum state we want to study. By recording the number of photons $n_1$ and $n_2$ in the beam splitter output channels, we get the information that is later used to reconstruct the state. Generally, the detectors have non-unit efficiencies, which we denote $\eta_1$ and $\eta_2$. We assume that each of the $m$ distributions is measured by $n$ representatives, so the total sample size is equal to $nm$.

In the strong-field approximation of local oscillator when $|\alpha_1| \to \infty$, registration of the reduced (differential) statistics $n_{12} = n_1 - n_2$ is equivalent to measure of the quadrature observable $X_\theta = \frac{1}{\sqrt{2}}\left(a \exp(-i\theta_1) + a^+ \exp(i\theta_1)\right)$, where $a$ and $a^\dagger$ are annihilation and creation operators, respectively. In this case, the following approximate equality is fulfilled: $X_\theta = n_{12}/(\sqrt{2}|\alpha|)$. The unitary operator of the beamsplitter is defined by the following expression, which depends on two parameters $\theta$ and $\varphi$: $U_{BS} = \exp\left(-\theta\left(\exp(-i\varphi)a_1^+ a_2 - \exp(i\varphi)a_1 a_2^+\right)\right)$. Here $a_1$, $a_2$, $a_1^\dagger$, $a_2^\dagger$ are annihilation and creation operators for the first and second modes, respectively. The standard 50/50 beamsplitter corresponds to the angles $\theta = \pi/4$, $\varphi = 0$.

The measurement of quadrature observables distributions $P(X_\theta)$ jointly with the inverse Radon transform leads to the reconstruction of the quantum state in the form of the Wigner function. Strictly speaking, the inverse Radon transform implicitly assumes that the measurement is made accurately and an infinite amount of data is collected, including an infinite number of "slices" of the phase space ($m \to \infty$) each including an infinite sample size ($n \to \infty$). The actual data, of course, does not satisfy these conditions. As a result, it often turns out that the state has been reconstructed with physically meaningless artifacts (for instance the density matrix has negative eigenvalues or contains negative numbers on the diagonal). Even if the artifacts are eliminated anyhow, the researcher has difficulty to draw the right conclusions about the accuracy, reliability and efficiency both for physical measurements and for mathematical reconstruction techniques. Thus, the historically important inverse Radon transform is not, after all, a rather rigorous and reliable method for the reconstruction of the quantum state (however, this is also true for all other methods of linear inversion).

Fortunately, in quantum tomography (QT) there have been developed a number of methods that allow one to achieve a reasonable estimate of the quantum states not only with an infinite, but also by a finite amount of experimental data. A detailed description of some of such methods is contained in references [7-11].

In this study, we rely on the so-called root approach to the considered problems, developed in our papers [12-16]. (QT) with the root approach suggests that we do not directly assess the density matrix of the quantum state $\rho$, but rather the square root of it $c = \sqrt{\rho}$ (hence the name of the method). The considered variable is the probability amplitude of the purified state. It is understood that the purified state is not uniquely defined, because if $U$ is an arbitrary unitary matrix, then $\sqrt{\rho}$ and $\sqrt{\rho}U$ are equivalent. This ambiguity, however, does not create any problems as the states correspond to the same density matrix $\rho$.

Note that the description of the density matrix directly as an object in the multidimensional space is a simple case only for a two-dimensional ($s = 2$) qubit state (in this case the geometry of the quantum state is given directly by Bloch sphere). However, with the growth of the dimension of the Hilbert space, the geometry of the state space becomes very intricate [17].

In our case, the quantum state is just an arbitrary complex matrix $c$ satisfying the normalization condition $Tr(cc^+) = 1$. Dimension of the matrix $c$ is $s \times r$, where $s$ is the dimension of the original Hilbert space, $r$ is the rank of the quantum state (the number of components in the mixture), $1 \leq r \leq s$. In fact, the purified quantum state "lives" in the extended Hilbert space of dimension $s\,r$. Simplicity of description of the space of possible states provides for simple iterative reconstruction procedures, as well as the possibility of describing the states of different rank starting from pure states ($r=1$) up to maximally mixed states ($r=s$). Another advantage of the approach is that it is possible to measure information about the parameters of the quantum state contained in respective protocols of quantum measurements.

There is no doubt that by taking measurements and actually destroying a certain set of the representatives of the quantum statistical ensemble, we obtain information about the quantum state. The question is how one measures the information quantitatively and how it can be used. Further, how does one determine whether the data is adequate to the studied physical object? How to translate the number of "clicks in the detectors" to the numerical values of the elements of the density matrix or state vector? How to evaluate the accuracy of the numerical values? How many "slices" corresponding to the different phases of the local oscillator should one take into account so that the measurements are complete? Finally, how many representatives in each slice have to be measured to assess the quantum state with a high predetermined accuracy?

The purpose of this paper is to get closer to the answers to these and some other questions. Note that similar questions have been partially studied in the papers [12-16], but for much more simple finite optical states that are based on polarization degrees of freedom. Otherwise, once we need to explore systems with an infinite number of degrees of freedom, we face the problem of the optimal truncation of the Hilbert space. On the one hand, such a reduced space should be wide enough to be able to accommodate a considered state and its possible fluctuation. On the other hand, a reduced space must be narrow enough to minimize the influence of the inevitable noise caused by the high dimensionality of the problem.

In this Letter, we limit ourselves by some examples of the analysis of single- and two-mode states, as well as a brief description of the methods that have been developed and used for such analysis. A detailed description of the methods and algorithms, as well as a detailed analysis of the multimode states, will be published somewhere else.

**Basis set of functions and examples of its applications**

In present paper we use a basis set of functions obtained from a set of Fock states $|k\rangle$ ($k=0,1,...$) through consistent application of displacement $D(\alpha) = \exp(\alpha a^+ - \alpha^* a)$ and squeezing $S(\xi) = \exp\left(\frac{1}{2}(\xi^* a^2 - \xi a^{+2})\right)$ operations: $|\alpha, \xi, k\rangle = S(\xi)D(\alpha)|k\rangle$, where $\alpha$ and $\xi$ are complex parameters.

The resulting set of functions $|\alpha, \xi, k\rangle$ as well as the initial set $|k\rangle$ are complete. Some special cases of the set are the ordinary *Fock states* $|k\rangle = |0,0,k\rangle$ (when $\alpha = 0$, $\xi = 0$), *coherent states* $|\alpha\rangle = |\alpha, 0, 0\rangle$ (when $\xi = 0$, $k = 0$), *squeezed vacuum states* $|\xi\rangle = |0, \xi, 0\rangle$ (when $\alpha = 0$, $k = 0$), *squeezed Fock states* $|\xi, k\rangle = |0, \xi, k\rangle$ (when $\alpha = 0$, $k \neq 0$), *squeezed coherent states* $|\alpha, \xi\rangle = |\alpha, \xi, 0\rangle$ (when $k = 0$) etc. In all these cases formed in the basis, the expansion will contain only single non-zero term. Thus, a wide range of important and widely used quantum states are represented in the simplest unified form.

Fig. 2 shows the results for a squeezed coherent state $|\alpha, \xi\rangle = |\alpha, \xi, 0\rangle$ with parameters $\alpha = 1 - i$, $\xi = 0.3\exp(i\pi/3)$. The two complex numbers define $\nu = 4$ real parameters - degrees of freedom of a quantum state. The parameters are estimated on the basis of numerical experiments, in accordance with Figure 1 with the following parameters of QT protocol: $|\alpha_1| = 2$, $m = 5$, $n = 500$.

Fig. 2a shows the Q-function for a given quantum state, Fig. 2b shows the distribution of fidelity $F$ between the unknown quantum state $\rho_0$ and its reconstruction $\rho$ for various cases. The considered value is determined by the following well-known formula $F = \left(Tr\sqrt{\rho_0^{1/2}\rho\rho_0^{1/2}}\right)^2$. The integral quality of the protocol is characterized by its efficiency, which is the ratio of the minimum possible average fidelity loss $\langle 1-F\rangle^*$ to the one actually observed: $e_P = \langle 1-F\rangle_{\min}/\langle 1-F\rangle$.

We see that analysis of the full statistics leads to a result that is close to ideal ($e_P = 0.93$). Such a protocol can be even closer to the ideal, if we increase the field of the oscillator $|\alpha_1|$ and the number of "slices" $m$ (eg, if $|\alpha_1| = 4$ then $e_P = 0.95$). We also see that the accuracy of QT is substantially reduced by the transition from full to reduced statistics, as well the transition from the ideal detector to an imperfect one. It is possible to understand the fundamental cause of this and make quantitative calculations of the corresponding based on the concept of information contained in the quantum measurements (see formula (2) below). For the five considered cases, the relevant complete information is: 10000, 10000, 6233, 4907, 3227.

Note that the case of complete statistics for an ideal detector comprises the maximum total amount of information. However, contrary to the ideal case this information is not quite uniformly distributed by degrees of freedom (2500, 2500, 2500, 2500 for the ideal case and 3334, 2907, 2093, 1666 for current case). Note that the information that corresponds to the components directly defines the variance of components, which, in turn, determine the distribution of fidelity.

In general, the reconstructed vector of a pure state is represented as a superposition:

$$|\psi\rangle = \sum_{k=0}^{s-1} c_k |\alpha,\xi,k\rangle \qquad (1)$$

Here $c_k$, $k = 0,1,...s-1$ is a set of complex amplitudes.

Equation (1) is exact for state $|\psi\rangle$ if this state is prepared in the framework of this basis. In a more general case the equation (1) is a finite- dimensional approximation of the state given in infinite-dimensional Hilbert space. Then it is assumed that $s$ is sufficiently large to approximate the state with the required accuracy.

As an example, Fig. 3 presents an analysis of superposition of the Fock state and coherent state: $N(c_\alpha|\alpha\rangle + c_n|n\rangle)$, where $c_\alpha$ and $c_n$ are corresponding amplitudes and $N$ is the normalization constant.

Fig. 3 shows a Q-function of the considered state with parameters, $\alpha = (1-i)/\sqrt{2}$, $n = 1$, $c_\alpha = c_n$. Figs. 3 b, c, d, e show the results of statistical reconstruction of the considered state. As a zero approximation the squeezed coherent state $|\alpha,\xi,0\rangle$ is chosen. However, unlike the example in Figure 2, in this case considering only one state is not sufficient, as it is necessary to consider the contribution of higher "harmonics" $|\alpha,\xi,k\rangle$. The development of an adequate model was carried out in two stages. As the first step forty basis functions $s = 40$ ($k = 0,1,...39$) were considered. At the second stage, the effective dimension of the Hilbert space has been reduced to $s = 9$. This was accomplished by moving from the initial basic functions to their most informative superpositions, which we call the principal components. QT was performed with the following parameters: $|\alpha_1| = 2$, $m = 7$, $n = 500$.

A total of 100 numerical experiments was carried out on the basis of complete statistics with ideal detectors. Fig. 3b shows that the results of experiments (histogram) are in good agreement with the theoretical distribution of fidelity (curve). The level of agreement by the chi - squared test is $\alpha_{crit} = 0.63$. The effectiveness of the considered protocol is $e_P = 0.95$.

---
* Sometimes this value is called *infidelity*

Fig. 3 c, d, e show the results illustrating the adequacy of the model. Note that the test of adequacy is formed by three mutually - complementary parts, namely, the agreement between exact theory and experiment (3 c), agreement between the developed model and experiment (3 d), as well as the agreement between the exact theory and developed a model (3 e).

It is worthy to note that the suggested approach also can be directly applied to the analysis of multiport systems. The following example illustrates the statistical reconstruction of the two-mode quantum state. We studied the following entangled state of two modes A and B:

$$|\psi_{AB}\rangle = \frac{|\alpha_A, k_1\rangle \otimes |\alpha_B, k_2\rangle + |\alpha_A, k_2\rangle \otimes |\alpha_B, k_1\rangle}{\sqrt{2}}$$

We assume here that the basis functions are not squeezed ($\xi = 0$).

The simulation results are presented in Figure 4. We have chosen the following parameters of the model: $\alpha_A = \frac{1-i}{2\sqrt{2}}$, $\alpha_B = \frac{1+i}{2\sqrt{2}}$, $k_1 = 1$, $k_2 = 2$.

In Figures (a) - (c) the considered two-mode state is illustrated graphically in the coordinate representation: the real part of the wave function (a), the imaginary part of the wave functions (b), density (c). Fig. 4 d shows a comparison of the results of numerical experiments (histogram) with the theoretical distribution for the loss of fidelityty (curve). 100 numerical experiments were performed. Their results are fully consistent with the theoretical prediction for the loss of fidelity. Significance level for the chi-square test for correspondence between theory and numerical experiments equals $\alpha_{crit} = 0.53$.

QT was carried out through a system of two beam splitters. The first inputs of the beamsplitters are fed by coherent modes of equal amplitudes of oscillators: $|\alpha_1^A| = |\alpha_1^B| = 1$. The second inputs of the respective beamsplitters are fed by modes A and B that are studied. The entanglement of the joint state of modes leads to a correlation of counts $(n_1^A, n_2^A)$ in channel A with the counts $(n_1^B, n_2^B)$ in channel B. Each of the oscillators provides a set of five different phases ($m_A = m_B = 5$). Similarly, the data has been collected from $m = m_A m_B = 25$ mutually complementary four-dimensional distributions $(n_1^A, n_2^A; n_1^B, n_2^B)$. The sample size in each such distribution was $n = 500$. The effectiveness of the protocol of quantum measurements was equal to $e_P = 0.90$.

**Reconstruction of the quantum state and the information contained in the quantum measurements**

The problem of statistical reconstruction of a quantum state is to restore in some sense the quantum state in the best way. In this paper, we start from a very general concept based on minimizing the distance between the empirical and reconstructed probability distributions [18].

It is remarkable that the considered approach can be directly generalized to the case of quantum measurement protocols. Let each line of protocol be identified by numbers $n_1, n_2$ and $\theta_j$, specifying the number of photons and the phase of the local oscillator and let $k_{n_1 n_2}^{\theta_j}$ be the corresponding number of observed events. Then, $\frac{1}{n} k_{n_1 n_2}^{\theta_j}$ is the frequency estimator of probability. Further, let $P_{n_1 n_2}^{\theta_j}$ be the estimator of the same probability that corresponds to the purified state $c$. Let the $\text{distance}\left(P^{\theta_j}_{n_1 n_2}, \frac{1}{n} k^{\theta_j}_{n_1 n_2}\right)$ be a functional defining the distance between these sets of probabilities. Our task is to find such pure state amplitude $c$, which will provide the minimum for the considered functional. It is remarkable that asymptotically when the amount of experimental data becomes large, the reconstruction results do not actually depend on which particular functional acts as the distance.

Note that in the case of mixed state of rank $r$ in the Hilbert space of dimension $s$ amplitude of the state $c$ can be pulled into a single column of length $rs$. To describe the accuracy of reconstruction it is convenient to transform the complex column state into a column of real numbers of length $2rs$. To do this, simply place the imaginary part of vector $c$ under the real part. Then, for quantitative description of the information contained in the quantum measurements, we can use the matrix of information, first introduced in [12]:

$$H = 2n \sum_j \frac{(\Lambda_j c)(\Lambda_j c)^+}{\langle c|\Lambda_j|c\rangle} \tag{2}$$

where $\Lambda_j$ are the measurement operators, defining the resolution of identity [19], separately for each phase of the local oscillator.

The matrix $H$ is a real symmetric matrix of dimension $2rs \times 2rs$. It is a measure of the information about the parameters of the quantum state, which is contained in the protocol of the quantum measurement.

Consider the vector $s_H$ composed of the eigenvalues of the matrix $H$ in decreasing order. In the case of complete protocol vector $s_H$ has $v_H = (2s-r)r$ non-zero strictly positive values, and the remaining $r^2$ values are exactly equal to zero. These eigenvalues define the information about the parameters of the quantum state. It appears that for unreduced ideal (pure) measuring the total information contained in the quantum measurements (the sum of the eigenvalues) is equal to $2nms$. The first maximum value that equals to $2nm$ is responsible for the normalization, and the other account for the accuracy of the parameter estimates of the quantum state. The number of degrees of freedom of a quantum state is $v = v_H - 1 = (2s-r)r - 1$. Each degree of freedom corresponds to its own value. All $v$ degrees of freedom correspond to the total information that is equal to $2nm(s-1)$. From the viewpoint of accuracy it is the best case when the information is uniformly distributed over all degrees of freedom. For the "bad" protocol, however, some degrees of freedom correspond to a small amount of information that leads to a drastic decrease in the accuracy of QT. Note that the transition to a reduced statistics and imperfect detectors reduces the information about the parameters of the state (in this case, if there is no loss of representatives of the quantum statistical ensemble, the information about the norm does not change).

The other $r^2$ of the $2rs$ real parameters are responsible for "non-physical" degrees of freedom - the arbitrary choice of global phase component of the mixture and ambiguity in the process of extraction of pure components of the density matrix. Each of these parameters corresponds to zero quantum information. In other words, the quantum measurement protocol contains no information about these parameters (and cannot contain according to the nature of quantum mechanics).

The information contained in the quantum measurement allows one to quantify strict reconstruction errors for each degree of freedom and for all states in general. The developed concept allows us to formulate the concept of the ideal distribution for the loss of fidelity, which minimizes the average loss and variance of fidelity. The approach allows for detailed description of a wide range of practically important measurement protocols with the reduced statistics, the non-unit effectiveness of the detector, multimode entanglement of states, etc. A quantitative characteristic of the quality of the protocol is its quantum efficiency $e_P$, which is a measure of the proximity of the protocol to the ideal one. The inequality $e_P \leq 1$ is in some sense a generalization of the Rao - Cramer inequality for the case of quantum measurements.

**Criterion of adequacy of QT experiment**

The so-called chi-squared parameter characterizes the degree of similarity between the observed frequencies of events and their expected values. It is calculated by the following formula

$$\chi^2_{ad} = \sum_{j=1}^{n_{bar}} \frac{\left(n_j^{\text{expected}} - n_j^{\text{observed}}\right)^2}{n_j^{\text{expected}}} \qquad (3)$$

Here $n_j^{\text{observed}}$ is the observed number of events, $n_j^{\text{expected}}$ is the expected number of events, $n_{bar}$ is the number of grouping intervals.

In the field of application of the chi-square test there are three different formulations of the problem. The first option is an assessment of the correspondence between the exact (a priori defined) theory and the experimental data. In this case, the $n_{bar}$ terms in (3) are bound by $m$ normalization conditions (by the number of studied mutual - complementary distributions), the number of remaining degrees of freedom is $v_{ad} = n_{bar} - m$. The second possibility is the assessment of the correspondence between the reconstructed model and experiment. In this case, besides the $m$ normalization conditions, there are $v = (2s - r)r - 1$ conditions imposed with accordance to the number of parameters of the quantum state. Then the number of the remaining degrees of freedom is $v_{ad} = n_{bar} - m - v$. Finally, a third possibility is an assessment of the correspondence between the exact theory and the reconstruction. In this case $v_{ad} = v$. The three possibilities are illustrated in Fig. 3 c, d, e.

It turns out that asymptotically (when the number of events is large) the parameter chi-square obeys the chi-square distribution with $v_{ad}$ degrees of freedom. This is a generalization to quantum protocol a well-known result dating back to Pearson and Fisher [20].

According to the well-known recommendations of statistics, the number of events in each grouping interval of data should be large enough (typically it is required that $n_j^{\text{expected}} \geq 5$).

It is crucial to understand that the criterion (3) is not applicable to all methods of estimation of quantum states, but only to the methods that provide optimal asymptotic estimates [18,20]. These methods include the considered method based on minimizing the distance between the collection of empirical and reconstructed distributions. To test the adequacy of the estimation one should also consider that the parameters of the quantum state are estimated by the grouped data (not by initial data). Finally, it is necessary that estimation of all parameters is supported by the presence of relevant information in the data to be measured.

Quantitatively, the adequacy of the QT experiment is characterized by the so-called critical level of significance $\alpha_{crit}$, which is the weight of a chi-square distribution with $v_{ad}$ degrees of freedom above the point $\chi^2_{ad}$. The model is considered an adequate one if $\alpha_{crit} > \alpha_0$, where $\alpha_0$ is a given significance level at the discretion of the experimenter (e.g. $\alpha_0 = 0.01; 0.05; 0.1$).

The Chi-square test is an effective tool for rejection of inadequate models. On the one hand, overly simplified models are inadequate, as the number of parameters is too small for a sufficiently accurate description of the data. On the other hand, overly complex models are also inadequate when the large number of parameters is not supported by the corresponding information of quantum measurements. As shown above as well as supported by similar numerical experiments, the chi-square test leads to identification of the quantum state with high accuracy.

**Conclusion**

This paper proposes a general approach which allows the experimenter to plan and implement effectively the research in the field of QT and utilize the available resources in the best way. This approach is based on the quantitative analysis of information about the parameters of the quantum state, which is contained in the applied quantum measurement protocol. Stringent information criteria of completeness and adequacy of quantum measurements are suggested and discussed, as well as a method of reconstruction, which provides accuracy close to that which is only achibable in principle for a given experimental constraints.

**Acknowledgments**
We would like to thank A.S. Holevo for many helpful discussions.
This work was supported in part by Russian Foundation of Basic Research (projects 14-02-00749, 13-07-00711), and by the Program of the Russian Academy of Sciences in fundamental research.

**References**

1. *Bohr N.* Discussion with Einstein on epistemological problems in atomic physics // in Schilp P.A. (editor), Albert Einstein, Philosopher-Scientist (Library of Living Philosophers, Evanston, Illinois, 1949), P.200-241.

2. *W. Pauli* General Principles of Quantum Mechanics Springer-Verlag Berlin Heidelberg New York 1980. 212 p.

3. *Horace P. Yuen, Vincent W. S. Chan* Noise in homodyne and heterodyne detection // Optics Letters. 1983. V. 8. № 3. p.177-179.

4. *Bonny L. Schumaker* Noise in homodyne detection // Optics Letters. 1984. V. 9. № 5. p.189-191.

5. *K. Vogel and H. Risken* Determination of quasiprobability distributions in terms of probability distributions for the rotated quadrature phase // Phys. Rev. A. 1989. V. 40. №5. P.2847-2849

6. *W.P. Schleich* Quantum Optics in Phase Space. Wiley-VCH 2001. 696 p.

7. *G. M. D'Ariano, M. G. A. Paris, and M. F. Sacchi,* Quantum Tomography, Advances in Imaging and Electron Physics, 128 205-308 (2003)- quant-ph/0302028;

8. *A. I. Lvovsky, M. G. Raymer* Continuous-variable optical quantum-state tomography // Rev. Mod. Phys. 2009. V.81. p.299-332.

9. *A. Ibort, V. I. Man'ko, G. Marmo, A Simoni and F. Ventriglia* An introduction to the tomographic picture of quantum mechanics // Phys. Scr. 79 (2009) 065013 (29pp)

10. *Paris, M., and J. Rehácek,* 2004, Eds., Quantum State Estimation Lect. Notes. Phys. Vol. 649. Springer Berlin Heidelberg.

11. Focus on Quantum Tomography. Editors Konrad Banaszek, Marcus Cramer and David Gross. Focus issue. New J. Phys. 2012-2013.

    http://iopscience.iop.org/1367-2630/page/Focus%20on%20Quantum%20Tomography

12. *Bogdanov Yu.I., Chekhova M.V., Krivitsky L.A., Kulik S.P., Penin A.N., Kwek L.C., Zhukov A.A., Oh C.H,. and Tey M.K.* Statistical Reconstruction of Qutrits // Phys. Rev. A. 2004. V.70. 042303

13. *Bogdanov Yu.I., Chekhova M.V., Kulik S.P., Maslennikov G.A., Zhukov A.A., Oh C.H. and Tey M.K.* Qutrit state engineering with biphotons // Phys. Rev. Lett. 2004. V.93. 230503.
14. *Yu. I. Bogdanov* Unified Statistical Method for Reconstructing Quantum States by Purification // Journal of Experimental and Theoretical Physics (JETP). 2009. Vol. 108. No. 6, pp. 928–935.
15. *Bogdanov Yu.I., Brida G, Genovese M., Kulik S.P., Moreva E.V., and Shurupov A.P.* Statistical Estimation of the Efficiency of Quantum State Tomography Protocols // Phys. Rev. Lett. 2010. V.105. 010404. 4p.;
16. *Yu. I. Bogdanov, G. Brida, I. D. Bukeev, M. Genovese, K. S. Kravtsov, S. P. Kulik, E. V. Moreva, A. A. Soloviev, A. P. Shurupov* Statistical Estimation of Quantum Tomography Protocols Quality // Phys. Rev. A. 2011. V.84. 042108. 19 p.
17. *I. Bengtsson, K. Zyczkowski* Geometry of Quantum States: An Introduction to Quantum Entanglement. Cambridge University Press. 2006. 417 p.
18. *Borovkov A.A.* Mathematical statistics. New York: Gordon and Breach. 1998. 570 p.
19. *Holevo A.S.* Statistical Structure of Quantum Theory. Springer. Lecture Notes in Phys., vol. 67, Springer-Verlag, 2001. 161 p.
20. *H. Cramer* Mathematical Methods of Statistics. Princeton University Press. 1946.

Figure captions

Fig.1 Scheme of homodyne measurements based on the beamsplitter

Figure 2 Study of a squeezed coherent state. Fig. 2 a: Q-function. Graphs in Fig. 2 b, from top to bottom: 1 - the ideal distribution ($e_P = 1$), 2 - a case of complete statistics for ideal detectors ($e_P = 0.93$), 3 - a case of complete statistics for non-ideal detectors with $\eta = 0.7$ ($e_P = 0.54$), 4 - the difference statistics $n_{12}$ for ideal detectors ($e_P = 0.41$), 5 - difference statistics for non-ideal detectors with $\eta = 0.7$ ($e_P = 0.25$).

Fig. 3. Analysis of the superposition of Fock state and coherent state. Q-function (a), the distribution of Fidelity loss (b), the criterion of the adequacy of the chi-square test for proximity measures between theory and experiment (c), between reconstruction and experiment (d) and between reconstruction and theory (e)

Figure 4. Reconstruction of the two-mode state. Two-mode psi-function in the coordinate representation: the real part (a), the imaginary part (b), density (c); Loss of fidelity (d)

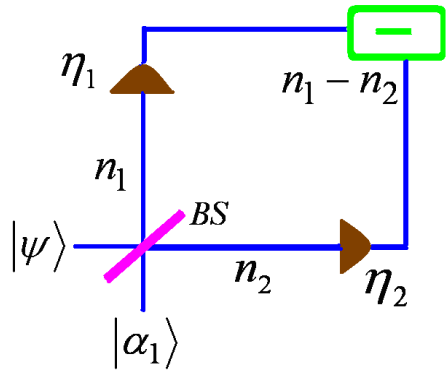

Fig.1

a)

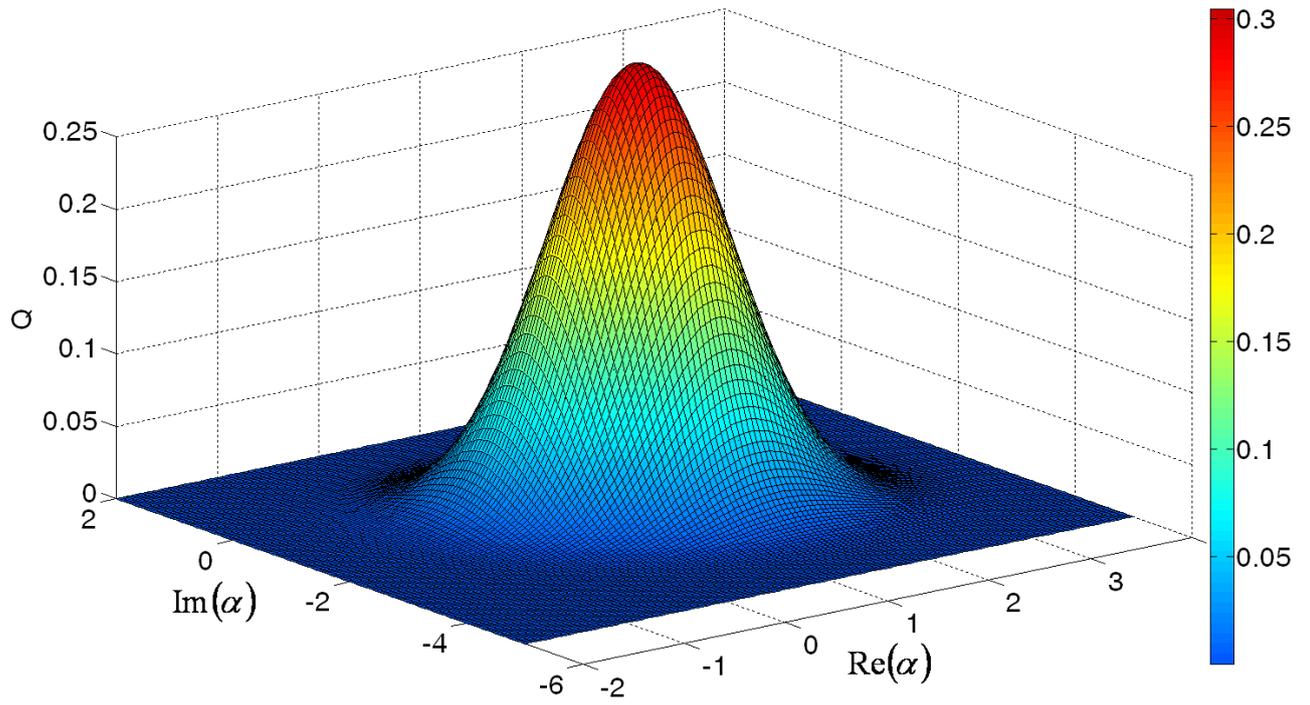

b)

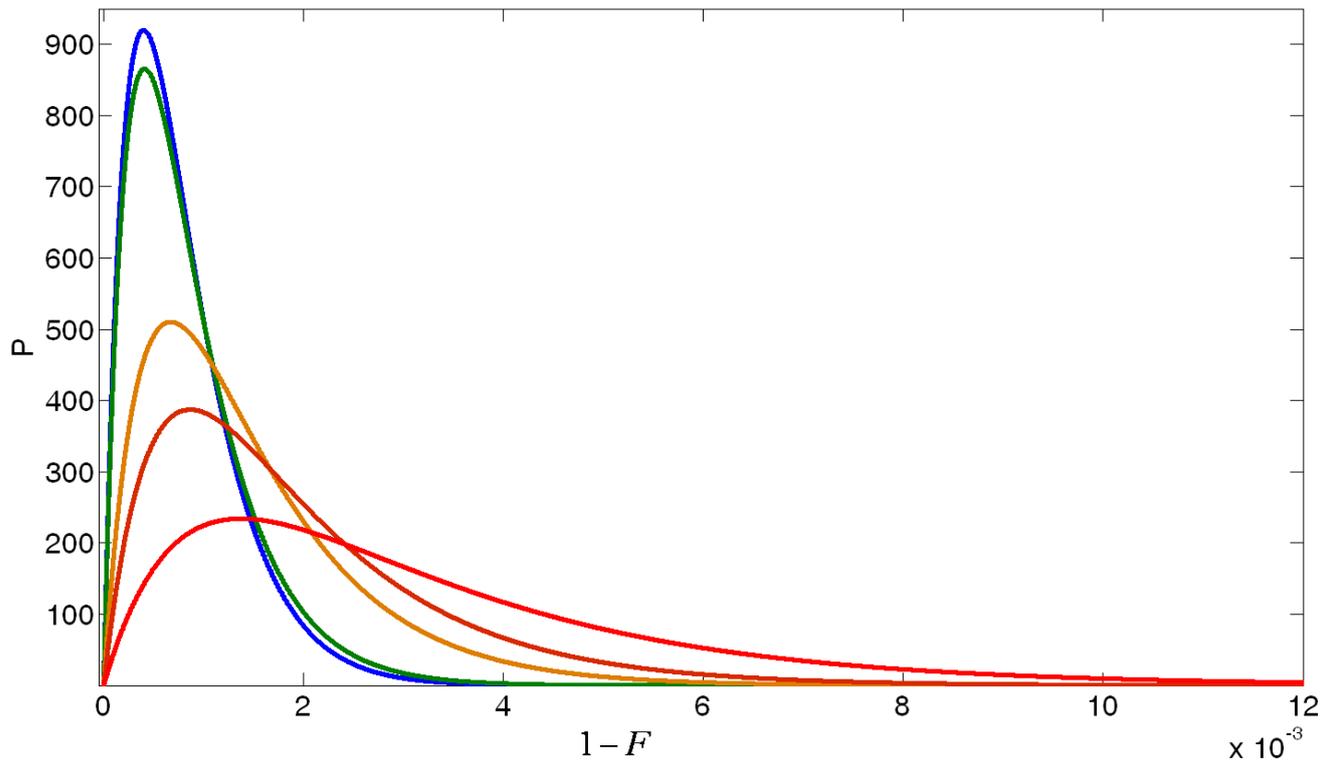

Fig.2

a)

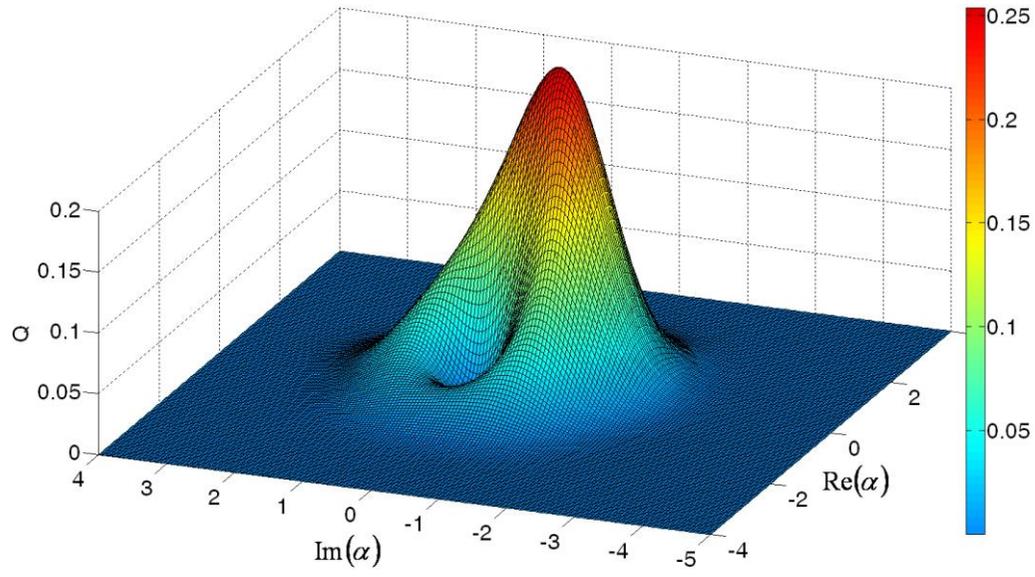

b)

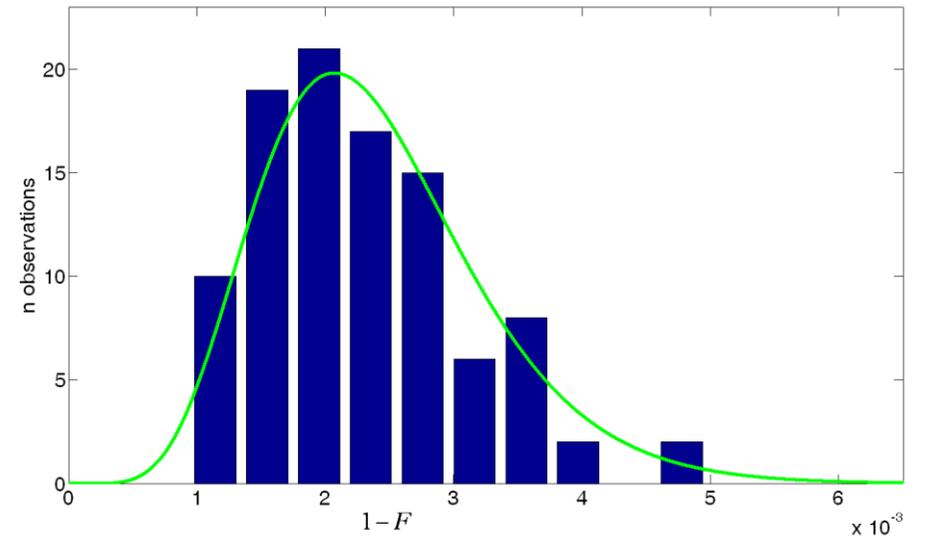

c)

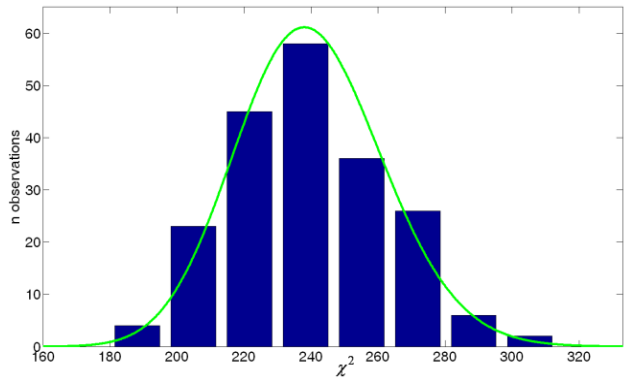

d)

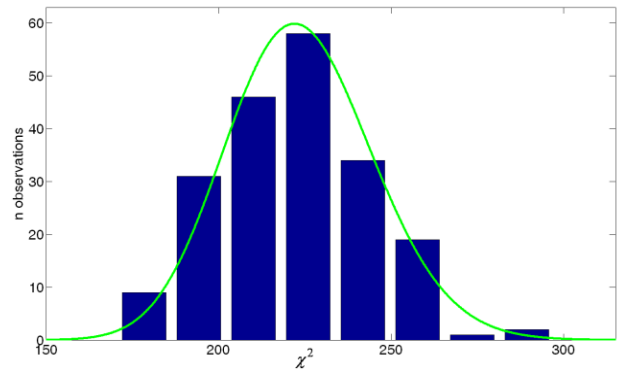

e)

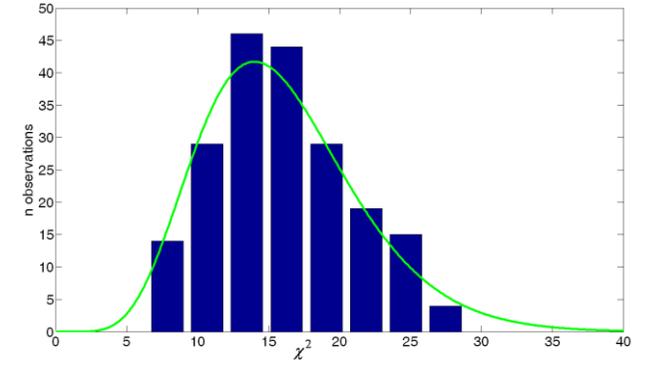

Fig.3

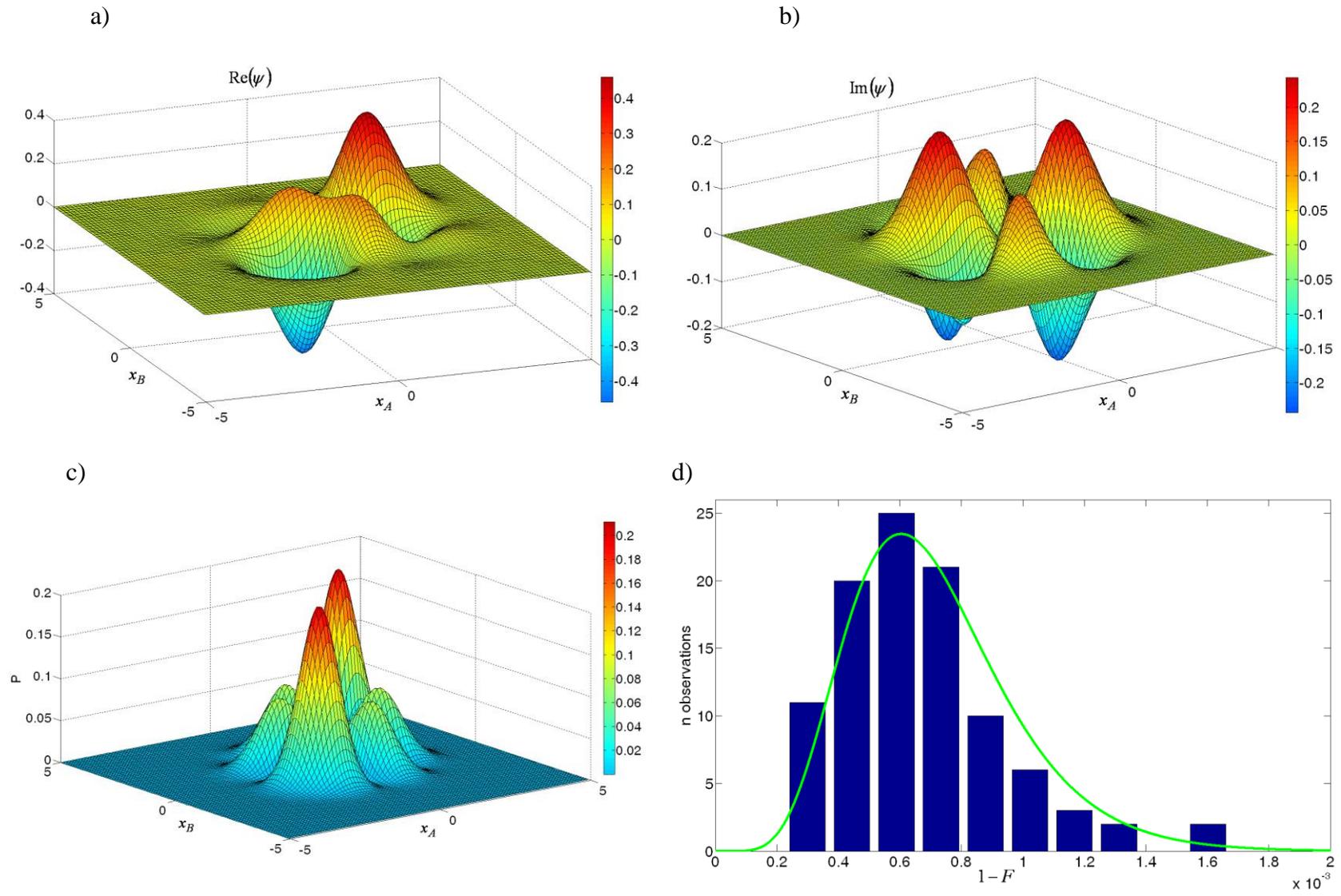

Fig.4

13